# Gauge bosons and fermions in SU(3)$_C$⊗SU(4)$_L$⊗U(1)$_X$ model with SU(2)$_H$⊗U(1)$_{AH}$ symmetry

Sutapa Sen

*Department of Physics ,Christ Church College ,Kanpur 208001,India*

*Email:sen.sutapa@gmail.com*

Abstract

We present a phenomenological study of neutral gauge bosons and fermions in an extended Standard model with SU(3)$_C$⊗SU(4)$_L$⊗U(1)$_X$ gauge symmetry .The model includes gauge bosons and fermions without exotic charges and is distinguished by the symmetry-breaking pattern SU(4)$_L$→SU(2)$_L$⊗SU(2)$_H$⊗U(1)$_{AH}$ This introduces an extended electroweak symmetry group SU(2)$_L$⊗SU(2)$_H$ at low energies We recover the fermion spectra of an anomaly-free three-family 3-4-1 model without exotic charges for U(1)$_X$ charge ,X = T$_{3R}$+ (B-L)/2 The interaction of physical neutral gauge bosons $Z', Z''$ and exotic fermions are presented along with their masses and mixing angle The electroweak constraints from oblique corrections on the model are also calculated.

PACS.12.60.Cn,12.15.Mm ,12.15Ff.

# 1. Introduction

An important challenge of the LHC is to look for signals of new physics beyond the Standard Model (SM)[1] which has several extensions with predictions of exotic fermions, gauge bosons and scalar bosons[2]. Among these, the class of left-right symmetric [L-R] models [3] have interesting predictions for neutrino masses through the see-saw scenario. In this work, we consider an extension of the left-right symmetric model with additional $SU(2)_H \otimes U(1)_{AH}$ symmetry which can be embedded in the $SU(3)_C \otimes SU(4)_L \otimes U(1)_X$ (3-4-1) gauge group. Recently, an extension of L-R symmetric model [4] has been proposed with additional global $U(1)_S$ symmetry with a generalized Lepton number [5]

The $SU(4) \otimes U(1)_X$ flavor symmetry has been considered in literature for both exotic [6] and ordinary electric charges[7] for exotic fermions and gauge bosons. This extension can answer the question of family replication with the requirement of number of families equal to the number of colors for anomaly cancellations. The Little Higgs Model (LHM) [8] with $SU(4)_L \otimes U(1)$ gauge symmetry has been recently considered in literature

In general, various 3-4-1 models can be classified by electric charge operator Q

$$Q = T_{3L} + \frac{b}{\sqrt{3}} T_{8L} + \frac{c}{\sqrt{6}} T_{15L} + X I_4 \qquad (1)$$

where $T_\alpha$, $\alpha = 3, 8, 15$ denotes diagonal generators of $SU(4)_L$ and X is the hypercharge operator for $U(1)_X$. $I_4$ is a 4×4 unit matrix. The values for X are fixed by anomaly cancellation of the fermion content of the three-family models. The condition of ordinary electric charges for fermions and gauge bosons for 3-4-1 model restrict three-family models to only of four different types with possible choices for parameters b and c [7] as b = 1, c = -2; b = -1, c = 2 ; b = c = 1.and b = 1, c = -1 .The hypercharge of the SM embedded in $SU(4)_L \otimes U(1)_X$ is $\frac{Y}{2} = \frac{b}{\sqrt{3}} T_{8L} + \frac{c}{\sqrt{6}} T_{15L} + X I_4$ .After symmetry breaking of 3-4-1 to SM, the gauge matching conditions give [9]

$$\tan \theta_W = \frac{g'}{g} ; \quad \frac{1}{g'^2} = \frac{b^2 + \frac{c^2}{2}}{3g^2} + \frac{1}{g_X^2} ; \quad \frac{g_X}{g} = \frac{\sin \theta_W (m_x)}{\sqrt{1 - \frac{3 + b^2 + \frac{c^2}{2}}{3} \sin^2 \theta_W (m_x)}} \qquad (2)$$

Here $g, g'$ are gauge coupling constants for $SU(2)_L$ and $U(1)_Y$, $g_X$ is the coupling constant for $U(1)_X$,

In this work, we consider the b = -1, c = 2 case with $\frac{g_X}{g} = \frac{\sin \theta_W (m_x)}{\sqrt{1 - 2\sin^2 \theta_W (m_x)}}$

It is interesting to note that the $U(1)_X$ charges for the fermion spectrum obtained after anomaly cancellations ( X) satisfy (B -2X) = L – 2$T_{3RV}$ where B, L denote baryon and lepton numbers respectively. $T_{3RV}$ is the third component of isotopic spin ( $T_{3R1}$+$T_{3R2}$) for a diagonal $SU(2)_{RV} \supset SU(2)_{1R} \otimes SU(2)_{2R}$ .Here SU(2)$_{R1}$ is the right-handed group for L-R symmetric model. SU(2)$_{R2}$ is an additional right handed group for exotic fermions. The values for X are obtained with distinct Baryon and Lepton numbers which is a special feature of this model,

$$X = T_{3R1} + T_{3R2} + \frac{(B-L)}{2} \tag{3}$$

The symmetry-breaking pattern is considered as in [9] with three stages. The first stage of symmetry-breaking $SU(4)_L \to SU(2)_L \otimes SU(2)_{HL} \otimes U(1)_{AH}$ is achieved by a 15-plet Higgs scalar boson S [9] The second and third stages are considered by introducing four Higgs scalar bosons, $\phi'\left(1,4,\frac{1}{2}\right), \chi'\left(1,4,-\frac{1}{2}\right), \phi\left(1,4,\frac{1}{2}\right), \chi\left(1,4,-\frac{1}{2}\right)$

$$SU(4)_L \otimes U(1)_X \xrightarrow{\langle S \rangle} SU(2)_{1L} \otimes SU(2)_{2HL} \otimes U(1)_{AH} \otimes U(1)_X$$

$$SU(2)_{1L} \otimes SU(2)_{2HL} \otimes U(1)_{AH} \otimes U(1)_X \xrightarrow{\langle \phi' \rangle, \langle \chi' \rangle} SU(2)_L \otimes U(1)_Y \otimes U(1)_{AH}$$

$$\xrightarrow{\langle \phi \rangle, \langle \chi \rangle} U(1)_{em} \otimes U(1)_{AH} \tag{4}$$

The diagonal generators of SU(4)$_L$ include T$_{3L}$, T$_{8L}$ and T$_{15L}$ which allow the combinations

$$A_H = \frac{1}{\sqrt{8}} Diag(1,1,-1,-1) = \sqrt{\frac{2}{3}}T_8 + \sqrt{\frac{1}{3}}T_{15} \; ; \; T_{3H} = -\sqrt{\frac{1}{3}}T_8 + \sqrt{\frac{2}{3}}T_{15} = \frac{1}{2}Diag(0,0,1,-1) \tag{5}$$

The charge operator Q is similar to that for the left-right symmetric model [3,4] with additional SU(2)$_{HL}$ left-handed group for exotic fermions instead of SU(2)$_R$

$$Q = T_{3L} + T_{3H} + XI_4. \tag{6}$$

The anomaly-free three families of fermions are listed in Table I along with (B-2X) for the fermion spectrum. The leptons $(v_e, e^-, N, E^-)_\alpha, \alpha = 1,2,3$ are assigned to the SU(4)$_L \otimes$U(1)$_X$ multiplet $L_\alpha \sim \left(1, 4, -\frac{1}{2}\right)$ which restrict these to only leptons (L = 1). The sterile neutrino is thus obtained in three generations along with a heavy lepton $E^-$. The model predicts a rich phenomenology for these leptons through interaction with new gauge bosons and four scalar bosons corresponding to SU(2)$_H \otimes$U(1)$_{AH}$ gauge symmetry

**Table I:** The fermion content, (B – 2X) and X charges for anomaly-free 3-4-1 model.

| Fermion | Content | Representation | (B -2X) | X |
|---|---|---|---|---|
| $Q_{iL}$, i = 1,2 | $(d_i, -u_i, D_i, U_i)_L$ | (3,4) | 0 | $\frac{1}{6}$ |
| $Q_{3L}$ | $(t, b, U, D)_L$ | (3,4*) | 0 | $\frac{1}{6}$ |
| $\bar{u}_{R\beta}, \beta = 1,2,3$ | $(\bar{u}, \bar{c}, \bar{t})_R$ | (3*,1) | 1 | $-\frac{2}{3}$ |
| $\bar{d}_{R\beta}, \beta = 1,2,3$ | $(\bar{d}, \bar{s}, \bar{b})_R$ | (3*,1) | -1 | $\frac{1}{3}$ |
| $\bar{U}_{R\beta}, \beta = 1,2,3$ | $(\bar{U}_{R1}, \bar{U}_{R2}, \bar{U}_{R3})$ | (3*,1) | 1 | $-\frac{2}{3}$ |
| $\bar{D}_{R\beta}, \beta = 1,2,3$ | $(\bar{D}_{R1}, \bar{D}_{R2}, \bar{D}_{R3})$ | (3*,1) | -1 | $\frac{1}{3}$ |
| $l_\beta, \beta = 1,2,3$ | $(v_\beta, e_\beta, N_\beta, E_\beta)_L$ | (1,4) | 1 | $-\frac{1}{2}$ |
| $(e^+_\beta, E^+_\beta)_R, \beta = 1,2,3$ | $(e^+, \mu^+, \tau^+)_R$ | (1,1) | -2 | 1 |
| | $(E_1^+, E_2^+, E_3^+)_R$ | (1,1) | -2 | 1 |

The work is organized as follows. In Section 2 we present the main features of the model and discuss its $SU(3)_L$ limit. Section 3 deals with the gauge boson sector with charged and neutral gauge bosons. We consider neutral currents and their mixing and obtain the electroweak constraints on the parameters of the model from oblique corrections. Section 4 is a short discussion on results and conclusions.

## 2. The 3-4-1 model without exotic charges

The 3-4-1 model without exotic charged particles [7] has been considered in literature for ( b=1,c = -2;b = -1,c = 2) [10]. For the second case, the electric charge operator

$Q = T_{3L} - \frac{1}{\sqrt{3}} T_{8L} + \frac{2}{\sqrt{6}} T_{15L} + X I_4$. The application of renormalization group equations discussed in [9] leads to ( $\sin^2\theta_W = \sin^2\theta_W(m_Z)$)

$$1 - 2\sin^2\theta_W - \frac{\alpha_X^{-1}(m_Z)}{\alpha^{-1}(m_Z)} = \frac{\alpha(m_Z)}{4\pi} \frac{44}{3} \ln \frac{m_X}{m_Z} \tag{7}$$

where $m_X$ is the unification scale, $\sin^2\theta_W(m_Z) = 0.2226$ and $\alpha(m_Z)^{-1} = 128.91$

The anomaly-free three families of fermions are listed in Table I. The X charges satisfy the condition of assumption (1) which relates (B-2X) to ($2T_{3RV} - L$) for the fermion spectrum.

The model includes four Higgs scalars with $X = \pm \frac{1}{2}$, $(\phi, \chi; T_{3R1} = \pm \frac{1}{2}); (\phi', \chi'; T_{3R2} = \pm \frac{1}{2})$

$$\chi \sim \left[1,4,-\frac{1}{2}\right] = \begin{pmatrix} \xi_\chi + i\varsigma_\chi \\ \chi_2^- \\ \chi_3^0 \\ \chi_4^- \end{pmatrix}; \quad \phi \sim \left(1,4,\frac{1}{2}\right) = \begin{pmatrix} \phi_1^+ \\ \xi_\phi + i\varsigma_\phi \\ \phi_3^+ \\ \phi_4^0 \end{pmatrix}$$

$$\chi' \sim \left[1,4,-\frac{1}{2}\right] = \begin{pmatrix} \chi_1'^0 \\ \chi_2'^- \\ \xi_{\chi'} + i\varsigma_{\chi'} \\ \chi_4'^- \end{pmatrix}, \quad \phi' \sim \left(1,4,\frac{1}{2}\right) = \begin{pmatrix} \phi_1'^+ \\ \phi_2'^0 \\ \phi_3'^+ \\ \xi_{\phi'} + i\varsigma_{\phi'} \end{pmatrix}$$

(8)

The neutral fields $\chi_3^0$, $\chi_1'^0$, $\phi_2'^0$ and $\phi_4^0$ are real while the remaining neutral fields are complex. The vacuum expectation values (VEV's) are aligned as

$$\langle\phi\rangle = \left(0,\frac{v}{\sqrt{2}},0,0\right)^T, \langle\chi\rangle = \left(\frac{u}{\sqrt{2}},0,0,0\right)^T; \langle\phi'\rangle = \left(0,0,0,\frac{V'}{\sqrt{2}}\right)^T, \langle\chi'\rangle = \left(0,0,\frac{V}{\sqrt{2}},0\right)^T$$

(9)

In addition, we consider a 15-plet Higgs scalar boson S transforming as (1,1,0) for first stage of symmetry-breaking $SU(4)_L \xrightarrow{\langle S \rangle} SU(2)_L \otimes SU(2)_H \otimes U(1)_{AH}$

$$\langle S \rangle = \frac{w}{2\sqrt{2}} diag(1,1,-1,-1)$$

(10)

**2.1 The SU(3)$_L$ electroweak symmetry limit of 3-4-1 model:** The model, in the SU(3)$_L$ limit, corresponds to the anomaly-free three-family 3-3-1 model [11] recently considered for phenomenology of exotic 2/3 charge quarks[12]. In this case, the electric charge operator

$$Q = T_{3L} - \frac{1}{\sqrt{3}} T_{8L} + X'; X' = \frac{2}{\sqrt{6}} T_{15L} + XI_4; X = T_{3R} + \frac{(B-L)}{2} \tag{11}$$

The Weinberg angle is now defined by

$$\tan \theta_W = \frac{g_Y}{g}, \quad \frac{1}{g_Y^2} = \frac{1}{3g^2} + \frac{1}{g'^2}; \quad \frac{g'^2}{g^2} = \frac{\sin^2 \theta_W (m_{X'})}{1 - \frac{4}{3}\sin^2 \theta_W (m_{X'})} \tag{12}$$

where $g, g'$ are now the coupling constants of SU(3)$_L$ and $U(1)_{X'}$

The scalar sector consists of three scalar triplets ρ (1, 3, 2/3 ), η(1,3,-1/3), χ(1,3,-1/3) which can be associated with φ(1,4,1/2), χ(1,4,-1/2) and χ′ (1.4.-1/2) scalar bosons in the 3-4-1 model. The study of the scalar sector in 3-3-1 case show that $\langle \xi_\rho \rangle, \langle \xi_\eta \rangle$ give mass to W$^\pm$ and Z bosons. The VEV's $\langle \xi_\rho \rangle = v, \langle \xi_\eta \rangle = u, \langle \xi_\chi \rangle = V$ where V >> v, u, and $\sqrt{u^2 + v^2} = v_{EW} = 246 GeV$

An important point of difference lies in the trilinear coupling $\varepsilon_{ijk} \chi_i \rho_j \eta_k$ which is introduced in the scalar potential [13] in 3-3-1 model. This is not present in the 3-4-1 case and thus affects the predictions for intermediate scalar masses in the 3-3-1 model. This is solved if we consider Higgs scalar $\Sigma*_{\alpha\beta}$ which belongs to the anti symmetric 6* representation of SU(4)$_L$. The symmetric 10$_S$ in $4 \otimes 4 \rightarrow 6^* + 10$ plays an important role in obtaining neutrino masses [9]. The role of $\Sigma_{\alpha\beta}$ is to introduce the trilinear coupling $\phi_\alpha \chi_\beta \Sigma_{\alpha\beta}$ in the scalar potential In the SU(3)$_L$ limit, this corresponds to 3* anti symmetric representation $\Phi_i^*$ of SU(3)$_L$ which should replace $\chi_i$ in the trilinear coupling as $\varepsilon_{ijk}\Phi_i\rho_j\eta_k$

The scalar sector is thus extended to add (6$_A$*+ 10s) SU(4)$_L$ multiplets which has interesting phenomenological implications for neutrinos and exotic fermions. This problem will be considered separately.

## 3. New Gauge bosons in 3-4-1 model.

The 24 gauge bosons associated with 3-4-1 model include W$_{i\mu}$ ( i =1,2,..15) in 15 –plet adjoint representation of SU(4)$_L$, one SU(4)$_L$ singlet B$_\mu$ associated with U(1)$_X$ and 8 gluons for SU(3)$_C$ color group The 15-plet representation of SU(4)$_L$ splits into mutiplets of

$SU(2)_L \otimes SU(2)_{HL} \otimes U(1)_{AH}$ as $15 = (3,1)_0 + (1,3)_0 + (1,1)_0 + (\bar{2},2)_{-2} + (2,\bar{2})_2$.

The covariant derivative for 4-plets is $D_\mu = (\partial\mu - igT_\alpha W_{\alpha\mu} - ig_X XI_4 B_\mu)$, where $\alpha = 1,2,..15$

$\mu = 1,2,3,4$. The coupling constants of SU(4)$_L$ and U(1)$_X$ are g and g$_X$ respectively.

$$T_\alpha W^\alpha{}_\mu = \frac{g}{\sqrt{2}} \begin{pmatrix} D^0_{1\mu} & W^+_\mu & Y^0_\mu & X'^+_\mu \\ W^-_\mu & D^0_{2\mu} & X^-_\mu & Y'^0_\mu \\ \bar{Y}^0_\mu & K^+_{1\mu} & D^0_{3\mu} & W'^+_\mu \\ X^-_\mu & \bar{Y}'^0_\mu & W'^-_\mu & D^0_{4\mu} \end{pmatrix} = \begin{pmatrix} (3,1)_0 & (\bar{2},2)_{-2} \\ (2,\bar{2})_2 & (1,3)_0 \end{pmatrix}$$
(13)

$$D_{1\mu}{}^0 = \frac{W_{3\mu}}{\sqrt{2}} + \frac{W_{8\mu}}{\sqrt{6}} + \frac{W_{15\mu}}{\sqrt{12}}; D_{2\mu}{}^0 = -\frac{W_{3\mu}}{\sqrt{2}} + \frac{W_{8\mu}}{\sqrt{6}} + \frac{W_{15\mu}}{\sqrt{12}}; D_{3\mu}{}^0 = -2\frac{W_{8\mu}}{\sqrt{6}} + \frac{W_{15\mu}}{\sqrt{12}}; D_{4\mu}{}^0 = -3\frac{W_{15\mu}}{\sqrt{12}}.$$
(14)

Introducing neutral gauge bosons $T_{3H\mu} = -\frac{1}{\sqrt{3}} W_{8L\mu} + \sqrt{\frac{2}{3}} W_{15L\mu}$; $A_{3H\mu} = \sqrt{\frac{2}{3}} W_{8L\mu} + \sqrt{\frac{1}{3}} W_{15L\mu}$

$$\sqrt{2} D^0_{1,2\mu} = \pm W_{3L\mu} + \frac{1}{\sqrt{2}} A_{3H\mu}; \sqrt{2} D^0_{3,4\mu} = \pm T_{3H\mu} - \frac{1}{\sqrt{2}} A_{3H\mu} \tag{15}$$

The charged gauge fields include

$$W^\pm_\mu = \frac{1}{\sqrt{2}} (W_1 \mp i W_2)_\mu ; X^\mp_\mu = \frac{1}{\sqrt{2}} (W_6 \mp i W_7)_\mu$$

$$X'^\pm_\mu = \frac{1}{\sqrt{2}} (W_9 \mp i W_{10})_\mu ; W'^\pm_\mu = \frac{1}{\sqrt{2}} (W_{13} \mp i W_{14})_\mu$$

$$Y^0_\mu = \frac{1}{\sqrt{2}} (W_4 - i W_5)_\mu ; Y'^0_\mu = \frac{1}{\sqrt{2}} (W_{11} - i W_{12})_\mu \tag{16}$$

After symmetry breaking with $\langle\phi\rangle + \langle\chi\rangle + \langle\phi'\rangle + \langle\chi'\rangle + \langle S\rangle$, and using covariant derivative $D_\mu$, we obtain the masses of charged gauge bosons

$$M^2_{W^\pm} = \frac{g^2}{4}(u^2 + v^2), M^2_{W'^\pm} = \frac{g^2}{4}(V^2 + V'^2);$$
$$M^2_{X^\pm} = \frac{g^2}{4}(V^2 + v^2), M^2_{X'^\pm} = \frac{g^2}{4}(u^2 + V'^2);$$
$$M^2_{Y^0} = \frac{g^2}{4}(V^2 + u^2), M^2_{Y'^0} = \frac{g^2}{4}(V'^2 + v^2). \tag{17}$$

For neutral gauge bosons, the combination $Z''_\mu = A_{3H\mu}$ is orthogonal to $T_{3H\mu}, W_{3L\mu}$.

The transformation between weak and mass basis is given by the rotation of the fields

$W_{3L\mu}, T_{3H\mu}, B_\mu$ into photon $A_\mu$, $Z_\mu$ and $Z'_\mu$ fields. Here $\tan\theta_P = t = \dfrac{\tan\theta_W}{\sqrt{1-\tan^2\theta_W}}$.

$$\begin{pmatrix} A_\mu \\ Z_\mu \\ Z'_\mu \end{pmatrix} = \begin{pmatrix} s_W & c_W s_P & c_W c_P \\ -c_W & s_W s_P & s_W c_P \\ 0 & -c_P & s_P \end{pmatrix} \begin{pmatrix} W_{3L\mu} \\ T_{3H\mu} \\ B_\mu \end{pmatrix} \qquad 18)$$

The neutral gauge boson fields $\left(A_\mu, Z_\mu, Z'_\mu, Z''_\mu\right)$ are

$$A_\mu = \sin\theta_W W_{3L\mu} + \cos\theta_W \left(\tan\theta_W T_{3H\mu} + \sqrt{1-\tan^2\theta_W} B_\mu\right)$$

$$Z_\mu = -\cos\theta_W W_{3L\mu} + \sin\theta_W \left(\tan\theta_W T_{3H\mu} + \sqrt{1-\tan^2\theta_W} B_\mu\right)$$

$$Z'_\mu = -\sqrt{1-\tan^2\theta_W}\, T_{3H\mu} + \tan\theta_W B_\mu; Z''_\mu = A_{3H\mu} \qquad (19)$$

The covariant derivative $D^0_\mu$ includes the physical neutral gauge fields $\left(A_\mu, Z_\mu, Z'_\mu, Z''_\mu\right)$

$$D^0_\mu = \partial_\mu - ieQA_\mu - i\frac{g}{\cos\theta_W}\left(T_{3L} - Q\sin^2\theta_W\right)Z_\mu - ig\left(-\sqrt{1-\tan^2\theta_W}\, T_{3H} + \frac{\tan^2\theta_W}{\sqrt{1-\tan^2\theta_W}} XI_4\right)Z'_\mu - igA_{3H}Z''_\mu$$

$$g_Z = \frac{g}{\cos\theta_W}; g_{Z'} = g_X \tan\theta_W; \frac{g_{Z'}}{g_Z} = t\sin\theta_W.$$

(20)

The squared mass matrix for neutral gauge bosons is obtained for the case of (1) equal VEV's

for SU(2)$_H$ Higgs scalar doublets, and a large VEV for singlet scalar S, with the hierarchy

$$w \gg (V = V') \gg u, v$$

$$M^2{}_N = \frac{g^2}{4c_W^2} \begin{pmatrix} (u^2+v^2) & -(u^2+v^2)ts_W & \frac{1}{\sqrt{2}}(u^2-v^2)c_W \\ -(u^2+v^2)ts_W & (u^2+v^2)t^2s_W^2 + \frac{2V^2c_W^2}{(1-t^2{}_W)} & \frac{1}{\sqrt{2}}(v^2-u^2)ts_W c_W \\ \frac{1}{\sqrt{2}}(u^2-v^2)c_W & \frac{1}{\sqrt{2}}(v^2-u^2)ts_W c_W & \frac{1}{2}(u^2+v^2+w^2+2V^2)c_W^2 \end{pmatrix} \quad (21)$$

For simplifying the mixing between gauge bosons we assume u = v for SM Higgs [14] in which case the $Z''$ gauge boson decouples from the other two,

$$M^2{}_N = \begin{pmatrix} M_Z^2 & -M_Z^2 ts_W & 0 \\ -M_Z^2 ts_W & \frac{g^2}{4}\left((u^2+v^2)t^2t^2{}_W + \frac{2V^2}{(1-t^2{}_W)}\right) & 0 \\ 0 & 0 & \frac{g^2}{8}(u^2+v^2+w^2+2V^2) \end{pmatrix} \quad (22)$$

The $Z''_\mu$ gauge boson acquires a heavy mass due to a large value of VEV $\langle S \rangle = w$. This distinguishes the model from other 3-4-1 cases [7, 10, 14] which predict $M^2{}_{Z''} \sim V^2$. Defining the squared mass terms

$$M_Z^2 = \frac{g^2}{4c_W^2}(u^2+v^2); M_{Z'}^2 = \frac{g^2}{4}\left\{(u^2+v^2)t^2t^2{}_W + \frac{2V^2}{(1-t^2{}_W)}\right\};$$

$$M_{Z''}^2 = \frac{g^2}{8}(u^2+v^2+w^2+2V^2); M_{ZZ'}^2 = -M_Z^2 ts_W. \quad (23)$$

For experimental comparison, we consider the general case [15] for a class of $Z'$ models with $M_{Z'} \gg M_Z$. The residual 2x2 $Z$-$Z'$ mixing mass matrix can be written as [15]

$$M_N^2 = M_Z^2 \begin{pmatrix} 1 & -ts_W \\ -ts_W & M_{Z'}^2/M_Z^2 \end{pmatrix} \tag{24}$$

We assume $M_{Z'}^2/M_Z^2 \gg 1$. The mixing angle $\phi$ between $Z$ and $Z'$ is introduced by defining physical neutral gauge bosons $Z_1$, $Z_2$ where

$$Z_{1\mu} = Z_\mu \cos\phi + Z'_\mu \sin\phi; Z_{2\mu} = Z_\mu \sin\phi - Z'_\mu \cos\phi$$

$$\tan 2\phi = \frac{2M_{ZZ'}^2}{(M_{Z'}^2 - M_Z^2)} . \phi = - t\sin\theta_W M_Z^2/M_{Z'}^2 \tag{25}$$

$$M_{Z_1,Z_2}^2 = \frac{1}{2}\left\{M_{Z'}^2 + M_Z^2 \pm \sqrt{(M_{Z'}^2 + M_Z^2)^2 - 4(M_Z^2 M_{Z'}^2 - \phi^2 M_{Z'}^4)}\right\} \tag{26}$$

The leading correction to Z boson mass is $\delta M_Z^2 = -\phi^2 M_{Z'}^2 = -t^2 \sin\theta_W^2 \frac{M_Z^4}{M_{Z'}^2}$. This leads to the masses for $Z_1$ and $Z_2$, $M_{Z_1}^2 = M_Z^2 - \delta M_Z^2; M_{Z_2}^2 = M_Z^2 + \delta M_Z^2$.

## 3.1 Neutral Currents

The interaction Lagrangian for $(Z_\mu, Z'_\mu, Z''_\mu)$ vector bosons and fermions is written as

$$L_{NC} = -\frac{g}{2c_W}\overline{f}\gamma^\mu(g_V - g_A\gamma^5)fZ_\mu - \frac{g}{2}\overline{f}\gamma^\mu(g'_V - g'_A\gamma^5)fZ'_\mu - g\overline{f}\gamma^\mu(g''_V - g''_A\gamma_5)fZ''_\mu \tag{27}$$

For SM gauge boson $Z_\mu$, $g_V = (T_{3L} - 2Qs^2_W), g_A = T_{3L}$. There are no couplings to exotic chiral fields except through $Z - Z'$ mixing. For $Z'_\mu$, $g_V' = \left(-T_{3H}\sqrt{1-t_W^2} + 2t_W Xt\right); g_A' = -T_{3H}\sqrt{1-t_W^2}$.

For $Z''_\mu$, $g_V'' = A_{3H}, g_A'' = A_{3H} = \frac{1}{2\sqrt{2}} Diag(1,1,-1,-1)$ The $Z_1 \to \bar{f}f, Z_2 \to \bar{f}f$ couplings are listed in terms of $g_{V1}$, $g_{V2}$, $g_{A1}$ and $g_{A2}$ where $\phi$ is the mixing angle.

$$g_{V1} = g_V + g_V'\phi, g_{A1} = g_A + g_A'\phi; g_{V2} = g_V\phi - g_V'; g_{A2} = g_A\phi - g_A'. \tag{28}$$

The SM fermions couple with $Z'_\mu$ universally through the vector coupling $g_V'$ by $2tt_W X$ term. The $Z''$ gauge boson couples uniformly to both SM and exotic fermions. The additional parameter w in the model can increase the mass above 1 TeV which affect the possibility of finding $Z''$ at LHC.

**Table II :** $Z_1 \to \bar{f}f, Z_2 \to \bar{f}f$ couplings to SM fermions

| Fermion | $g_{V1}$ | $g_{V2}$ | $g_{A1}$ | $g_{A2}$ |
|---|---|---|---|---|
| $t_L$ | $\left(\frac{1}{2}-\frac{4}{3}s_W^2\right)+\frac{tt_W}{3}\phi$ | $\left(\frac{1}{2}-\frac{4}{3}s_W^2\right)\phi-\frac{tt_W}{3}$ | $\frac{1}{2}$ | $\frac{1}{2}\phi$ |
| $b_L$ | $\left(-\frac{1}{2}+\frac{2}{3}s_W^2\right)+\frac{tt_W}{3}\phi$ | $\left(-\frac{1}{2}+\frac{2}{3}s_W^2\right)\phi-\frac{tt_W}{3}$ | $-\frac{1}{2}$ | $-\frac{1}{2}\phi$ |
| ($s_L,d_L$) | $\left(\frac{1}{2}+\frac{2}{3}s_W^2\right)+\frac{tt_W}{3}\phi$ | $\left(\frac{1}{2}+\frac{2}{3}s_W^2\right)-\frac{tt_W}{3}\phi$ | $\frac{1}{2}$ | $\frac{1}{2}\phi$ |
| ($c_L,u_L$) | $\left(-\frac{1}{2}-\frac{4}{3}s_W^2\right)+\frac{tt_W}{3}\phi$ | $\left(-\frac{1}{2}-\frac{4}{3}s_W^2\right)\phi-\frac{tt_W}{3}$ | $-\frac{1}{2}$ | $-\frac{1}{2}\phi$ |
| $e_i$ | $\left(-\frac{1}{2}+2s_W^2\right)-tt_W\phi$ | $\left(-\frac{1}{2}+2s_W^2\right)\phi+tt_W$ | $-\frac{1}{2}$ | $-\frac{1}{2}\phi$ |
| $\nu_i$ | $\frac{1}{2} - tt_W\phi$ | $\frac{1}{2}\phi - tt_W$ | $\frac{1}{2}$ | $\frac{1}{2}\phi$ |

**Table III:** $Z_1 \to \bar{f}f, Z_2 \to \bar{f}f$ couplings to exotic fermions

| Exotic fermion | $g_{V1}$ | $g_{V2}$ | $g_{A1}$ | $g_{A2}$ |
|---|---|---|---|---|
| $U_L$ | $\left(-\frac{1}{2}\sqrt{1-t_W^2} + \frac{tt_W}{3}\right)\phi$ | $\left(\frac{1}{2}\sqrt{1-t_W^2} - \frac{tt_W}{3}\right)$ | $-\frac{1}{2}\sqrt{1-t_W^2}\phi$ | $\frac{1}{2}\sqrt{1-t_W^2}$ |
| $D_L$ | $\left(\frac{1}{2}\sqrt{1-t_W^2} + \frac{tt_W}{3}\right)\phi$ | $\left(-\frac{1}{2}\sqrt{1-t_W^2} - \frac{tt_W}{3}\right)$ | $\frac{1}{2}\sqrt{1-t_W^2}\phi$ | $-\frac{1}{2}\sqrt{1-t_W^2}$ |
| $U_1, U_2$ | $\left(\frac{1}{2}\sqrt{1-t_W^2} + \frac{tt_W}{3}\right)\phi$ | $\left(-\frac{1}{2}\sqrt{1-t_W^2} - \frac{tt_W}{3}\right)$ | $\frac{1}{2}\sqrt{1-t_W^2}\phi$ | $-\frac{1}{2}\sqrt{1-t_W^2}$ |
| $D_1, D_2$ | $\left(-\frac{1}{2}\sqrt{1-t_W^2} + \frac{tt_W}{3}\right)\phi$ | $\left(\frac{1}{2}\sqrt{1-t_W^2} - \frac{tt_W}{3}\right)$ | $-\frac{1}{2}\sqrt{1-t_W^2}\phi$ | $\frac{1}{2}\sqrt{1-t_W^2}$ |
| $N_i, i=1,2,3$ | $\left(-\frac{1}{2}\sqrt{1-t_W^2} - tt_W\right)\phi$ | $\left(\frac{1}{2}\sqrt{1-t_W^2} + tt_W\right)$ | $-\frac{1}{2}\sqrt{1-t_W^2}\phi$ | $\frac{1}{2}\sqrt{1-t_W^2}$ |
| $E_i, i=1,2,3$ | $\left(\frac{1}{2}\sqrt{1-t_W^2} - tt_W\right)\phi$ | $\left(-\frac{1}{2}\sqrt{1-t_W^2} - tt_W\right)$ | $\frac{1}{2}\sqrt{1-t_W^2}\phi$ | $-\frac{1}{2}\sqrt{1-t_W^2}$ |

Table II lists the $Z_1$ and $Z_2$ couplings to SM fermions and in Table III we list these couplings to exotic chiral fields. The shift in $Z\bar{f}f$ coupling for chiral field f is $\varepsilon_f = \frac{g}{\cos\theta_W} t \sin\theta_W X_f \phi$

## 3.2 Electroweak Constraints on parameters of model

We now obtain constraints on the mass of $Z'$ and parameter V from the following [15]

(1) The effect of shift in Z boson mass on radiative corrections [16] can be related to oblique parameter T (Peskin -Takeuchi parameter) [17] to give oblique correction [15]

$$\alpha T = -\frac{\delta M_Z^2}{M_Z^2} = \frac{t^2 \sin^2\theta_W M_Z^2}{M_{Z'}^2} = \tan^4\theta_W \frac{v^2}{2V^2} = g_X^4 \frac{1}{(g^2+g_X^2)^2} \cdot \frac{v^2}{2V^2}. \qquad (29)$$

This is similar to the result derived in LHM [see S. Nam in ref.[8]].

From recent experimental data[18], $\alpha_S(M_Z) = 1/127.925$ and $\sin^2\theta_W = 0.2226$

Since $t = \frac{\sin\theta_W}{\sqrt{1-\sin^2\theta_W}}$ and v= $\sqrt{u^2+v^2} = 246 GeV$, we obtain a constraint on the parameter V.

(a) At 95% CL for central value of Higgs mass $M_H$ =117 GeV, T < 0.06 gives V > 2.2697 TeV and the mass of Z' gauge boson as $M_{Z'}$ = 1.2477 TeV. The Z-Z' mixing angle $\phi$ = 0.0014.

(b) For T $\leq$ 0.12, eqn (29) gives V $\geq$ 1.604 TeV and $M_Z'$ = 890.6065 GeV. The

Z-Z' mixing angle $\phi$ = 0.0028

Electroweak constraints on model from effective weak charge $Q_W$ in atomic parity violation

is obtained for mixing angle $\phi$ by the relation

$$Q_W = -2\{(2Z+N)C_{1u}+(Z+2N)C_{1d}\} \qquad (30)$$

where $C_{1u}, C_{1d}$ are given by $C_{1q} = 2g_Z g_{A1}^e (g_{V1}^q + \frac{tt_W}{3}\phi)$

This result depends on $\tan\theta_W$ and $g_Z$ which receive contributions from oblique corrections. The discrepancy between the SM and recent experimental data, $\Delta Q'_W = 0.45 \pm 0.48$ for cesium atom.

For T < 0.06 ( 0.12), $\phi$ = 0.0014 (0.0028) the model predicts $\Delta Q'_W$ = 0.09 (0.187) without oblique correction to $\sin\theta_W$ Thus the range of V and $M_Z^/$ considered are consistent with electroweak constraints .In Table IV we present a summary of our results.

The extended gauge group $SU(4)_L$ breaks to give an additional $SU(2)_{HL}$ symmetry which gives $(1,3)_0$ $W'_\mu$ gauge boson .These gauge bosons are predicted in simplest extensions of SM [19].In the present model these gauge bosons do not mix with charged gauge bosons of the SM $(3,1)_0$ and can couple only to exotic fermion fields . From eqn.(17), the mass $M_{W'}$ is predicted at a scale V .The additional gauge bosons $(\bar{2},2)_{-2}, (2,\bar{2})_2$ can couple SM fermions to exotic fermion fields and have a lower mass than $M_{W'}$.These have interesting one-loop contributions to $Z'$ decays .

The flavor-changing neutral currents (FCNC) also constrain the model. The symmetry-breaking pattern considered for 3-4-1 model is [10]

$$SU(4)_L \otimes U(1)_X \supset SU(3)_L \otimes U(1)_z \supset SU(2)_L \otimes U(1)_Y$$

The FCNC parameters have been recently used to constrain the mass of exotic quark U at 3-3-1 level [12].As discussed in Section II, this is the SU(3) limit of the present model and predicts U mass in TeV range. A detailed analysis in the 3-4-1 case would through light on the D, $U_1$ and $U_2$ fermion masses. The lepton sector consists of three generations of sterile neutrino and heavy lepton E⁻ .These leptons have been predicted by some models [21] models and have been suggested at 2-14 TeV energies.

## 4. Conclusions

In this work, we have presented an analysis of $SU(4)_L \otimes U(1)_X$ model without exotic electric charges. The symmetry is broken according to $SU(4)_L \rightarrow SU(2)_L \otimes SU(2)_{HL} \otimes U(1)_{AH}$

The electric charge operator Q consists of parameters b = -1 and c = 2, which leads to

Q = $T_{3L}$+$T_{3HL}$+ $XI_4$.This is an extended L-R symmetry $SU(2)_L \otimes SU(2)_{HL} \otimes U(1)_X \otimes U(1)_{AH}$ .The X charge obtained after anomaly-cancellations in all 3-4-1 models now corresponds to the hypercharge Y of L-R model $X = T_{3R1} + T_{3R2} + \frac{(B-L)}{2}$ which assigns Baryon and Lepton numbers to exotic fermions . The U(1)$_{AH}$ is an additional factor in the extended group with a new gauge boson $Z''$ which has mass determined by VEV w of a scalar S. Such bosons, predicted by many extensions of SM present important challenges at LHC [20].The mass of neutral gauge boson $Z'_\mu$ and the Z-$Z^/$ mixing have been obtained at the approximation $V = V' \Box\ u = v$ .The constraints on new physics are used from oblique corrections and effective weak charge in atomic parity violation to limit the lower bound of the parameter V and lower limit of $M_{Z'}$ .These occur at energies available to LHC , at around (0.89-1.25) Tev for $M_{Z'}$ .The SU(2)$_{HL}$ group consists only of exotic fermions which are singlets of SM, while their right-handed singlet antifermions belong to $T_{3R2} = \pm\ ½$.

The 3-4-1 model without exotic charges (b = -1,c = 2) also has a three-family anomaly-free

3-3-1 model embedded in it [12] .The phenomenology of exotic fermions will be studied in further work.